\documentclass[conference]{IEEEtran}
\pdfoutput=1
\IEEEoverridecommandlockouts
\usepackage{cite}
\usepackage{amsmath,amssymb,amsfonts}
\usepackage{algorithmic}
\usepackage{subcaption}
\usepackage{multirow,bigdelim}
\usepackage{booktabs}
\usepackage{graphicx}
\usepackage{textcomp}
\usepackage{xcolor}
\def\BibTeX{{\rm B\kern-.05em{\sc i\kern-.025em b}\kern-.08em
    T\kern-.1667em\lower.7ex\hbox{E}\kern-.125emX}}

\begin{document}

\title{Exploiting Temporal Dependencies for Cross-modal Music Piece Identification
\\
\thanks{The research is supported by the European Union under the EU's Horizon 2020 research and innovation programme, Marie Sk\l{}odowska-Curie grant agreement No. 765068.
The LIT AI Lab is supported by the Federal State of
Upper Austria.}
}

\author{\IEEEauthorblockN{Lu\'is Carvalho$^1$, Gerhard Widmer$^{1,2}$}
\IEEEauthorblockA{\textit{$^1$Institute of Computational Perception
$\&$ $^2$LIT Artificial Intelligence Lab} \\
\textit{Johannes Kepler University Linz, Austria}
}}

\maketitle

\begin{abstract}

This paper addresses the problem of cross-modal musical piece identification and
retrieval: finding the appropriate recording(s) from a database 
given a sheet music query, and vice versa, working directly with audio and scanned
sheet music images.
The fundamental approach to this \cite{DorferSVKW18_CCALayer_IJMIR} is to learn a
cross-modal embedding space with a suitable similarity structure for audio and sheet
image snippets, using a deep neural network, and identifying candidate pieces by
cross-modal near neighbour search in this space. 
However, this method is oblivious of temporal aspects of music. In this paper, we
introduce two strategies that address this shortcoming.
First, we present a strategy that aligns sequences of embeddings
learned from sheet music scans and audio snippets.
A series of experiments on whole piece and fragment-level retrieval
on 24 hours worth of classical piano recordings demonstrates significant
improvement.
Second, we show that the retrieval can be further improved by 
introducing an attention mechanism to the embedding learning model that reduces the
effects of tempo variations in music.
To conclude, we assess the scalability of our method and discuss potential measures to make it suitable for truly large-scale applications.
\end{abstract}


\section{Introduction}
\label{sec:intro}

Large amounts of music-related contents are available nowadays in the
digital realm, in diverse forms, from studio and live
audio recordings to scanned sheet music images and video clips.
Making such heterogeneous collections searchable and
explorable in a content-based way requires efficient techniques for 
cross-linking between items of different modalities.
In \textit{cross-modality document retrieval}, 
we have a collection of items of a certain modality (e.g., music recordings)
and wish to retrieve relevant documents from this
by querying with items of a different modality (e.g., scores) -- either
entire documents or fragments thereof.


In this paper we address the problem of \textit{score-based piece identification}.
Our goal is to perform this task in \textit{both search directions}, which means
finding a score from a collection given an audio query and, inversely,
retrieving an appropriate audio performance given a sheet music input.
We attempt to solve this problem in its most extreme setup, in the absence of
any metadata or machine-readable information: we work directly with raw 
material, that is, audio recordings and digitised images of scanned sheet 
music.




Previous work \cite{DorferHAFW18_MSMD_TISMIR} has shown how to perform 
audio-sheet music piece identification with a two-stage procedure, by 
retrieving short snippets of music and then generating a ranked list 
by counting the number of retrieved snippets per piece.
The audio-to-score correspondences were obtained by learning a cross-modal 
embedding space for both audio and score snippets by means of a 
deep neural network \cite{DorferSVKW18_CCALayer_IJMIR}.
Despite encouraging results, a number of challenges have remained open.
Most importantly, the counting-based strategy entirely neglects the 
inherent temporal dependencies between music snippets, 
both on the score and audio side.
And second, the network architecture is not designed 
to account for (possibly large)
tempo variations in music, where varying speed greatly affects the 
amount of audio/visual content in fixed-size snippets, making the
approach rather brittle.
%

%
Our central contribution is a musically more meaningful identification 
procedure that exploits the strong temporal relations between 
consecutive audio and score snippets, aiming for more robust and 
accurate identification.
The basic idea will be to compute a matching function by aligning 
subsequent snippets of a query and corresponding retrieved items,
both projected  onto a learned embedding space, using a dynamic time warping 
(DTW) \cite{Mueller15_FMP_SPRINGER} algorithm whose cost is defined over
pairwise distances in this multi-modal embedding space.


A first experiment (Section \ref{subsec:exp1}) evaluates our alignment-based 
procedure against the voting approach in \cite{DorferHAFW18_MSMD_TISMIR}, 
revealing a significant boost in performance, which
indicates the positive impact of introducing an alignment method to this task.
This latter point is further corroborated by an experimental comparison
to an adapted version of the identification algorithm used in the popular
\textit{Shazam} system \cite{Wang03_Shazam_ISMIR}, which does not rely on alignment.
To make the snippet retrieval step more robust to extreme differences in 
tempo between score and audio excerpts, we then add a 
soft-attention mechanism (as described in \cite{BalkeDCAW19_ASR_TempoInv_ISMIR})
to the baseline architecture (Section \ref{subsec:exp2}), permitting
the network to decide the appropriate temporal 
context for a given query snippet.
Experimental results show that the compound effect of addressing the
two aforementioned problems amounts to $ 280 \%$ and $ 80 \%$ improvement 
over the baseline method for audio-to-score and score-to-audio retrieval, 
respectively.

The above experiments aligned entire scores and pieces. To get an understanding of how our system behaves under a more realistic usage scenario, we
conduct a set of fragment-level retrieval 
experiments, systematically varying the lengths of the queries (Section \ref{subsec:exp3}), giving more insight into the relation between query length
and identification accuracy.
Lastly we investigate how our system behaves when increasing the dataset size (Section \ref{subsec:exp4}), and
discuss opportunities to make it suitable for very large music collections.
All the experiments in this work are conducted with commercial recordings 
of several hundred complex classical piano pieces (24 hours worth of audio), and their respective sheet music scans.
%

\section{System description}
\label{sec:description}

%
We first explain how to connect audio to sheet music images by
learning an embedding space, then introduce our identification procedure that aligns snippets of musical content.
The main premise is that the documents of both modalities -- audio recordings and score images -- have been cut, in a pre-processing step, into a set of fixed-size segments (\textit{snippets}); these are the items for which we wish to learn a joint embedding space
that should place related snippets of both modalities in close proximity, to permit distance-based retrieval \cite{DorferHAFW18_MSMD_TISMIR}.

%

\subsection{Learning Audio-to-score Relations}
\label{subsec:baseline}

\begin{figure}[t]
\centering
\centerline{\includegraphics[width=0.65\columnwidth]{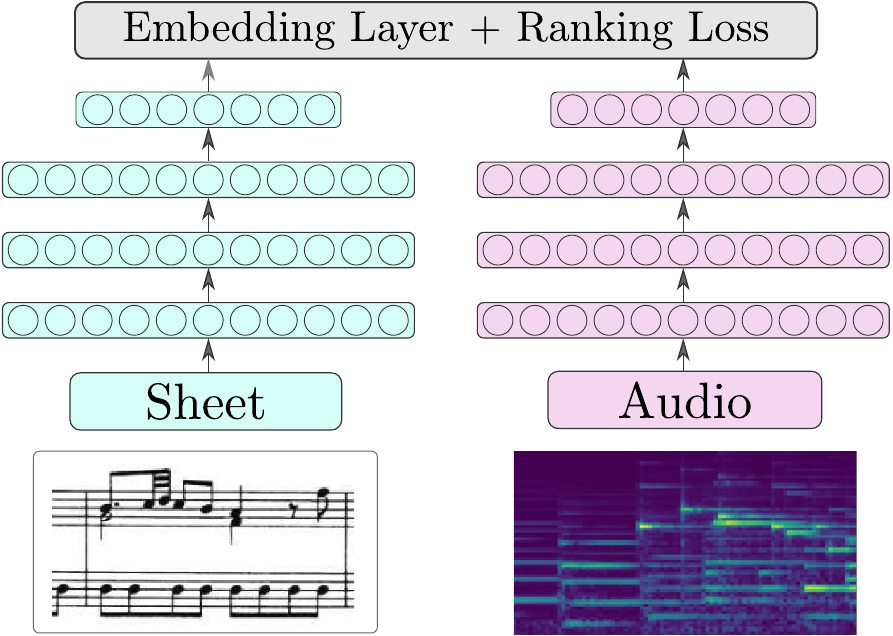}}
\caption{Architecture of cross-modal embedding network~\cite{DorferHAFW18_MSMD_TISMIR}.}
\label{fig:network}
\end{figure}

To learn a cross-modal score/audio embedding space, we employ a deep
neural network model depicted in 
Fig.\ref{fig:network}.
%
%
It consists of two convolutional pathways, each 
responsible for embedding one of the music modalities.
The canonically correlated (CCA) embedding layer ensures that the outputs of 
the two pathways are projected into a shared 32-dimensional 
space~\cite{DorferSVKW18_CCALayer_IJMIR}.
The network is trained by optimising a pairwise ranking
loss~\cite{KirosSZ14_VisualSemanticEmbeddings_arxiv}: the cosine 
distances between corresponding snippet pairs are minimised, while
distances between non-corresponding pairs are maximised.
This results in matching pairs being projected close to each other and
dissimilar ones falling apart.
To train our model, we use the MSMD dataset~\cite{DorferHAFW18_MSMD_TISMIR},
which contains over 300,000 score-audio snippet pairs 
from synthesised classical piano music.

One advantage of this approach is that pairs of audio and
sheet music snippets are required only during training.
At test time, each pathway can embed their corresponding snippets independently.
Thus, the network can operate in both retrieval directions: 
audio-to-score and score-to-audio.
%


\subsection{Piece Identification via Snippet Voting}
\label{subsec:piece_ident}

%
The retrieval task now consists in finding a corresponding audio recording when given a score scan, or the correct score when given a recording.
The basic method for cross-modal identification works as follows.
We outline the steps for the audio-to-score direction (the audio 
recording is the query), but emphasise that the opposite direction 
works analogously.

Let $ \mathcal{D} $ be a collection of
$ L $
images of sheet music pages,
and $ Q $ an audio query.
Each document $ D_i \in \mathcal{D} $ is processed by a system 
detector to automatically identify system coordinates 
in the score and then cut into a set of image snippets as the one 
shown in Fig.\ref{fig:network}.
The snippets are then embedded by passing them through
the score pathway of the trained network, resulting in a set of sheet music 
embeddings $ \{ y^i_1, y^i_2, ..., y^i_{M_i} \} $ for each piece.
Analogously, the audio query is segmented into short spectrogram
excerpts, which are embedded via the audio pathway of the model, resulting
in a set of audio embeddings $ \{ x_1, x_2, ..., x_N \} $ 
for the query.

Given this database of sheet music images 
and an audio query embedded to the same shared space,
a two-stage strategy for piece identification was employed  in~\cite{DorferHAFW18_MSMD_TISMIR} that generates a ranked list via snippet 
retrieval.
First for each audio snippet $ x_j $ of the query, its nearest neighbour 
from the database of all embedded image snippets is selected via cosine distance.
%
Each retrieved snippet then votes for the piece it originated from,
resulting in a ranked list of piece candidates.
%
From now on, this will be our \textit{baseline} method.

\subsection{Exploiting Temporal Dependencies}
\label{subsec:alignment}

The vote-based procedure completely ignores the temporal relationships 
between subsequent snippet queries, which are key in music.
Since both query and database items are now segmented and projected onto a shared 
space, a piece (score and audio recording) can be seen as a sequence
of snippet embeddings, with a distance metric defined in this
space, also between snippets of different modality.
Thus, we can use an \textit{alignment procedure} to test how well an audio query
(as a sequence of audio snippet embeddings) and a score (as a sequence of score snippet
embeddings) `fit' together, using the embedding space distance as a cost function.
A similar approach was adopted in~\cite{FremereyMKC08_AutomaticMapping_ISMIR}, 
where sheet image scans are converted into chroma features via an optical music 
recognition system and aligned to audio recordings.

In the following, we first assume that our query is always a full piece 
(recording or score). This will be relaxed later, in Section \ref{subsec:exp3}.
%
%
%
%
We formalise the matching procedure similarly to 
\cite{BalkeALM16_BarlowRetrieval_ICASSP}, but replace the Subsequence 
Dynamic Warping (SDTW) by its standard DTW algorithm, as we are aligning 
entire sequences.
The sequence of embedded snippets $ \{ y^i_1, y^i_2, ..., y^i_{M_i} \} $ 
of each piece $ D_i \in \mathcal{D} $ from the database\footnote{\label{note1}
For now, we simply go through the entire database of pieces and
run a DTW alignment on all of them. Obviously, one could first obtain a subset of $k$ top
candidates via voting-based retrieval, which would then be further checked and
re-ranked via DTW, but as DTW is so efficient -- especially given the low (snippet-level)
resolution of our sequences --, it turned out that our simple procedure, which does
not require near-neighbour search in embedding space, is basically equally
fast.}
is aligned  to the query sequence $ \{ x_1, x_2, ..., x_N \} $ via DTW, using the cosine 
distance as a cost function. The DTW alignment cost between 
query $ Q $ and piece $ D_i $ is regarded as the matching cost 
$ c_i = \mathrm{DTW} (Q, D_i) $.
Finally we generate a ranked list based on the matching cost of 
each piece to the query, with the best matching piece 
having the lowest alignment cost.

\section{Experiments}
\label{sec:experiments}


%

\begin{figure}[t!]

\begin{minipage}[b]{1.0\linewidth}
\centering
\centerline{\includegraphics[width=0.97\columnwidth]{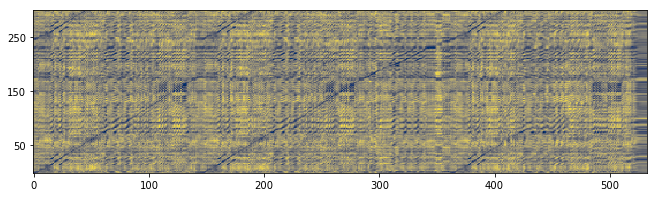}}
\centerline{(a) Mozart KV 280, 1st movement}\medskip
\end{minipage}

\begin{minipage}[b]{1.0\linewidth}
\centering
\centerline{\includegraphics[width=0.97\columnwidth]{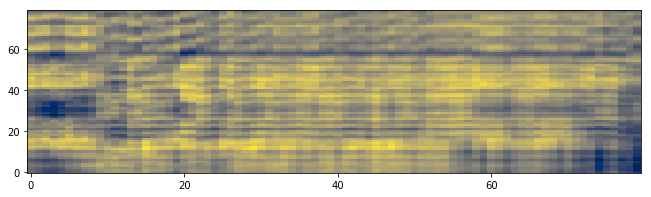}}
\centerline{(b) Chopin Prelude Op.28 No.22}\medskip
\end{minipage}

\caption{Distance matrices (x axis: audio; y axis: score) 
        of two cases where standard DTW fails: (a) presence of repeats; 
       (b) when the embedding projections are meaningless.}
\label{fig:bad_cases}

\end{figure}


%
We use a collection of $321$ commercial recordings of 
classical piano pieces and their corresponding sheet music scans, which 
consists of our private music collection plus scores obtained from the IMSLP
online library\footnote{https://imslp.org} and recordings retrieved from 
Youtube\footnote{https://www.youtube.com/}.
This amounts to over $24$ hours of recorded music and 1,696 
pages of scores, and comprises complex pieces such as Chopin Ballades
and Beethoven and Mozart piano sonatas.
The scores were manually scanned, are slightly noisy, and 
the score grid lines (staff lines, bars, etc) are not perfectly
aligned to the scanned document margins, bringing our retrieval scenario 
closer to real cases.

Moreover, manual inspections revealed that at least $84$ of the $321$ pieces 
($26 \%$) manifest some sort of structural mismatch 
between score and  audio recording
-- mainly due to repeats played in the audio,
but not written out in the score.\footnote{We currently do not have a reliable
method for identifying and correctly interpreting repeat signs, \textit{da capo / dal segno}
indications, etc.}
%
This is rather challenging, as the standard DTW algorithm 
cannot handle such structural differences.
To illustrate, we show in Fig.\ref{fig:bad_cases}(a) the distance 
matrix of a Mozart piece with repeated sections, which can be seen as the main
dark diagonal path jumps to the beginning of the score.
%


Embeddings are generated as in Section \ref{subsec:baseline}.
All score pages are first resized to 
a width of $835$ pixels, and $160 \times 200 $-pixel snippets are 
sequentially cut from them with a hop size of $50$ pixels.
We replaced the CNN-based system detector 
\cite{GallegoCZ17_StaffLineRemoval_Autoenc_ESA} used in 
\cite{DorferHAFW18_MSMD_TISMIR} with an open source software 
\cite{BitteurH13_AUDIVERIS}, resulting in $32 \%$ more systems 
accurately detected.
Audio recordings are transformed into $92$-bin log-frequency spectrograms, and excerpts of roughly $2$ seconds of music are segmented with a hop size of approximately half a second,
resulting in audio snippets with 
dimension $ 92 \times 42 $ (bins $ \times $ frames).

We evaluate our piece identification procedure on 
both query directions: audio-to-score and score-to-audio; from now on we
refer to these tasks as A2S and S2A, respectively.
For evaluation we calculate standard  metrics for document-level retrieval
from the resulting ranked lists: 
different recalls $R @ K$, looking at the top $K$ matches;
mean reciprocal rank (MRR; higher is better) and median rank (lower is better).

\begin{table}[t!]
 \centering
 \begin{subtable}{\columnwidth}
 \centering
 \scalebox{0.95}{
 \begin{tabular}{lcccccc}
 \toprule
 \textbf{Method} & \bfseries R@1 & \bfseries R@5 & \bfseries R@10 & \bfseries MRR & \bfseries MR\\
 \midrule
 Baseline~\cite{DorferHAFW18_MSMD_TISMIR}& 43 (0.13) & 93 (0.29) & 135 (0.42) & 0.23 & 15 \\
 Baseline+Att  & 93 (0.29) & 188 (0.59) & 231 (0.72) & 0.42 & 4 \\
 \midrule
 DTW & 195 (0.61) & 257 (0.80) & 279 (0.87) & 0.69 & 1 \\
 DTW+Att & 266 (0.83) & 292 (0.91) & 303 (0.94) & 0.87 & 1 \\
 \midrule
 Shazam~\cite{Wang03_Shazam_ISMIR} & 129 (0.40) & 173 (0.54) & 193 (0.60) & 0.47 & 4 \\
 Shazam~\cite{Wang03_Shazam_ISMIR}+Att & 154 (0.48) & 189 (0.59) & 210 (0.65) & 0.54 & 2 \\
 \bottomrule
\end{tabular}}%
\caption{Audio-to-score (A2S) piece identification results.}
\label{tab:exp_a2s}
\end{subtable}

\vspace{5pt}
\begin{subtable}{\columnwidth}
 \centering
 \scalebox{0.95}{
 \begin{tabular}{lcccccc}
 \toprule
 \textbf{Method} & \bfseries R@1 & \bfseries R@5 & \bfseries R@10 & \bfseries MRR & \bfseries MR\\
 \midrule
 Baseline~\cite{DorferHAFW18_MSMD_TISMIR} & 115 (0.36) & 183 (0.57) & 209 (0.65) & 0.46 & 4 \\
 Baseline+Att  & 185 (0.58) & 242 (0.75) & 262 (0.82) & 0.66 & 1 \\
 \midrule
 DTW & 225 (0.70) & 266 (0.83) & 292 (0.91) & 0.76 & 1 \\
 DTW+Att &  247 (0.77) & 288 (0.90) & 298 (0.93) & 0.83 & 1 \\
 \midrule
 Shazam~\cite{Wang03_Shazam_ISMIR} & 129 (0.40) & 172 (0.54) & 187 (0.58) & 0.46 & 4 \\
 Shazam~\cite{Wang03_Shazam_ISMIR}+Att & 140 (0.44) & 174 (0.54) & 191 (0.60) & 0.49 & 3 \\
 \bottomrule
\end{tabular}}%
\caption{Score-to-audio (S2A) piece identification results.}\label{tab:exp_s2a}
\end{subtable}
 \caption{Piece identification results for both query directions.
R@k: Recall@k, MRR: Mean Rec.~Rank, MR: Median Rank.}
\label{tab:exp_results}
\end{table}

\subsection{Experiment 1: Baseline vs.~Alignment}
\label{subsec:exp1}

In the first experiments, we compare our 
alignment-based matching procedure to the baseline method.
Additionally, we adapted the \textit{Shazam} search method 
\cite{Wang03_Shazam_ISMIR} to our task by 
using the embeddings as the spectral fingerprints from the original 
formulation, in order to compare our findings with an efficient (non-alignment-based)
benchmark framework.
%
We define a search space $ \mathcal{D} $ with all $321$ pieces in one modality
and a query set $ \mathcal{Q} $ consisting of their respective counterparts.
Then we query each piece $ Q_i \in \mathcal{Q} $ in the search collection 
$ \mathcal{D} $ and compute the aforementioned metrics.
The results are summarised in Table~\ref{tab:exp_results} (methods without ``+Att").

Overall we see that the DTW-based strategy performs better, by a large margin, 
than the baseline and also the Shazam method 
for all evaluation metrics, in both search directions.
The baseline retrieves around $13 \%$ of the queries (43 pieces) correctly as the 
best match ($R @ 1$) in the A2S task, whereas the DTW method correctly 
retrieves $61 \%$ (195 pieces).
Similar improvements can be seen in the S2A direction.
At higher $k$, the recall values reach $90\%$ levels. Also noteworthy
is the median rank of 1, indicating that in the majority of cases, the correct
answer indeed appears at the top of the list.
%
%
%
%
Generally, the Shazam search method performed better than the baseline, but not 
as well as the alignment-based strategy.

A few observations can be derived from this.
First, the considerable improvements of our method indicate that the 
learned representations can support meaningful DTW synchronisation paths.
The baseline vote-based strategy relies solely on snippet-wise distance 
computations, which may not exhibit the expected projection characteristics 
(see~\ref{subsec:baseline}) for all excerpts, whereas the DTW-based method can
overcome local projection mismatches by computing an overall alignment 
path.
Second, we noticed a trend that if a piece creates problems
in one query direction, it also performs poorly in the opposite 
direction.
Manual inspection revealed that such pieces failed to generate meaningful
embedding vectors, therefore producing a poor alignment path for the 
audio-score piece pair.
To illustrate, Figure~\ref{fig:bad_cases}(b) shows the distance 
matrix of Chopin's Prelude No. 22, which gives the worst 
retrieval results (ranked at $28$ and $179$ for A2S and S2A, respectively).
No evident alignment path is visible from the matrix, and the DTW 
algorithm appears to have found less costly warping paths by synchronising the 
query to other pieces.

\subsection{Experiment 2: Attention Mechanism}
\label{subsec:exp2}

In a second experiment, we modify the snippet embedding model by introducing a 
soft-attention mechanism \cite{BalkeDCAW19_ASR_TempoInv_ISMIR} to
the audio input.
Since in the original approach the snippet dimensions are fixed, tempo 
variations will inevitably affect the amount of musical content that audio 
excerpts will accommodate.
This leads to discrepancies between what the 
network sees during training, and the test data.
%
%
As a solution, \cite{BalkeDCAW19_ASR_TempoInv_ISMIR}
proposes to increase the audio field of view and let the network decide on the 
appropriate temporal context for the given audio snippet, by adding an attention 
branch to the audio pathway of the network.
%
%
When comparing different audio snippet lengths,  
we achieve the best identification results using an 
audio context of $4$ seconds ($84$ frames).
%
The same context window is applied to all methods indicated by \textit{+Att} in 
Table~\ref{tab:exp_results}.


We observe additional improvement on both A2S and S2A tasks, for both baseline and DTW-based methods.
%
%
%
In both directions, now at least $93 \%$ of the sought pieces are correctly returned 
among the top 10 ranks.
More generally, the experiment supports several interesting observations.
%
First we note that learning better representations, which produced superior 
audio-to-score snippet retrieval results \cite{BalkeDCAW19_ASR_TempoInv_ISMIR}, 
also leads to better piece identification.
By simply adding the attention model to the baseline vote-based strategy, the 
MRR improves by approximately $83 \%$ and $43 \%$ for tasks A2S and S2A, 
respectively.
Moreover, we observe that, while in \cite{BalkeDCAW19_ASR_TempoInv_ISMIR} the 
attention models were evaluated only for the audio-to-score snippet retrieval 
direction, the updated models also show positive impact on sheet-to-audio 
direction tasks, for all methods.
Modifying the network architecture and adopting an alignment-based matching 
strategy revealed the most substantial improvement over the baseline:
$280 \%$ and $80 \%$ better for the MRR, on the A2S and S2A task, respectively.

\begin{table}[t!]
 \centering
 \begin{subtable}{\columnwidth}
 \centering
 \scalebox{0.95}{
 \begin{tabular}{lccccccccc}
 \toprule
 \textbf{Length (s)} & \bfseries 10 & \bfseries 20 & \bfseries 30 & \bfseries 40 & \bfseries 50 & \bfseries 60 & \bfseries 70 & \bfseries 80 \\
 \midrule
 MRR & 0.41 & 0.51 & 0.56 & 0.59 & 0.61 & 0.62 & 0.63 & 0.64 \\
 \bottomrule
\end{tabular}}%
\caption{Audio-to-score direction.}
\label{tab:exp3_a2s}
\end{subtable}

\vspace{5pt}
\begin{subtable}{\columnwidth}
 \centering
 \scalebox{0.95}{
 \begin{tabular}{lccccccccc}
 \toprule
 \textbf{Systems} & \bfseries 1 & \bfseries 2 & \bfseries 3 & \bfseries 4 
 & \bfseries 5 & \bfseries 6 & \bfseries 7 & \bfseries 8 \\
 \midrule
 MRR & 0.45 & 0.54 & 0.59 & 0.63 & 0.63 & 0.63 & 0.63 & 0.64 \\
 \bottomrule
\end{tabular}}%
\caption{Score-to-audio direction.}
\label{tab:exp3_s2a}
\end{subtable}
 \caption{Fragment-level retrieval results by varying the length of the query
 (MRR = Mean Reciprocal Rank).}
\label{tab:exp3_results}
\end{table}

\subsection{Experiment 3: Fragment-level Retrieval}
\label{subsec:exp3}

In this set of experiments, we test our method on incomplete queries,
modelling a realistic scenario where one may not have the entire piece 
for a search, but a fragment of an unknown audio recording or 
an unlabelled page 
of sheet music, and wishes to identify its originating piece.
A similar study on MIDI-to-score retrieval was recently described in 
\cite{Tsai20_LinkingLakhtoIMSLP_Bootleg_ICASSP}.

First, we replace the DTW by its variant SDTW \cite{Mueller15_FMP_SPRINGER}, 
to permit synchronisation of short sequences to longer ones.
We follow the pipeline described in \cite{BalkeALM16_BarlowRetrieval_ICASSP}.
For the audio queries, we vary the fragment context from $10$ to $80$ seconds, 
in increments of $10$.
On the score side, we vary the query sizes in system units (rows in the score), 
from one to eight.
For each fragment size, 1,500 queries are randomly selected from our database 
and the search is performed and evaluated as before.
%

The main observation (see Table~\ref{tab:exp3_results}) is the upward trend of the MRR values 
as queries get longer.
The biggest jumps in performance were obtained when going from 
$10$ to $20$ seconds in the A2S retrieval direction, and 
from one to two systems on the S2A task.
Interestingly, in both directions, the 
performance increase rate tends to decrease as the queries get bigger.
A possible explanation is the number of pieces  
with structural differences in our collection (see above):
a longer query is more likely to contain such deviations, 
and increasing its length appears to be not as meaningful as expected.

\subsection{Experiment 4: Scalability}
\label{subsec:exp4}

The last set of experiments evaluates the scalability of our method.
As digitisation of music content rises and online music/score
archives can reach the order of tens of thousands of items, retrieval systems 
are expected to scale to larger collections.
Our current music collection is limited in size; however, we can use it to experimentally investigate how well our method scaled to our current data 
volume.
%
%

We randomly select smaller subsets of the main collection and perform short queries 
in both directions, as in Section~\ref{subsec:exp3}.
For each subset we randomly 
select 1,000 fixed-size query fragments, the sizes being 50 seconds and four 
lines in the score for the audio and score queries, respectively.
%
Then we measure the final MRR value and the average search time per query.
We repeat this procedure 10 times for each subset size and 
use the average of the results, similarly
to~\cite{Tsai20_LinkingLakhtoIMSLP_Bootleg_ICASSP}.

\begin{figure}[t!]

\begin{minipage}[b]{1.0\linewidth}
\centering
\centerline{\includegraphics[width=0.97\columnwidth]{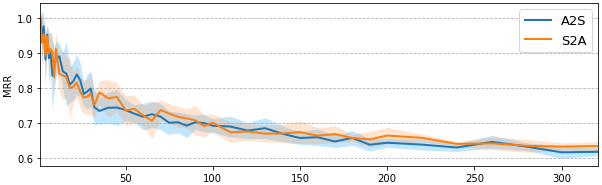}}
\centerline{(a) Mean reciprocal rank across different dataset sizes}\medskip
\end{minipage}

\begin{minipage}[b]{1.0\linewidth}
\centering
\centerline{\includegraphics[width=0.97\columnwidth]{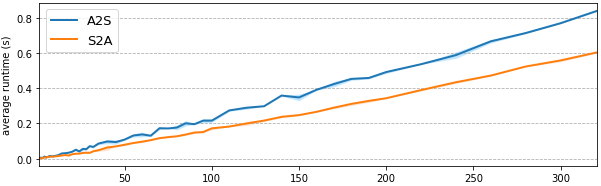}}
\centerline{(b) Average search time across different dataset sizes}\medskip
\end{minipage}

\caption{(a) MRR and (b) average search time for many dataset sizes, which are 
indicated by the horizontal axis
(given in terms of number of entire pieces. Note that, on average, a single piece is 
represented by around 523 audio excerpts and 422 score snippets; a dataset size of 
300 pieces thus corresponds to roughly 160,000 excerpts and 130,000 snippets)
}
\label{fig:scalable}

\end{figure}

Figure~\ref{fig:scalable} shows the MRR and average search time for 
both search directions as the dataset size increases.
Regarding the MRR, we consider that the system scaled moderately up to our 
current collection size. For instance, the MRR value dropped less than 0.1 as the 
dataset size increased from 100 to 321, for both A2S and S2A.

%

Regarding the average search time, as expected the system scaled roughly
linearly, since the time complexity of DTW is
$O(MN)$~\cite{Mueller15_FMP_SPRINGER}, with $M$ and $N$ the size of the database 
and the query length (fixed), respectively.
%
It is clear that without any efficiency improvement measures, this would not be
practicable for huge music repositories. There are fairly obvious solutions to this.
%
One way to improve the numbers in absolute terms is to adopt the re-ranking 
strategy described in Footnote~\ref{note1} and accelerate the voting process via 
faster indexing techniques.
In~\cite{SchnitzerFW09_FastRetrieval_ISMIR} we showed how to achieve speedups
of up to 40 times compared to an exhaustive linear scan,
by using a filter-and-refine
approach, which made it possible to efficiently answer nearest neighbour queries in a collection of 2.5 million songs. An analogous approach, adapted to our distance measure, would also be applicable in our present system.
%


\section{Conclusion and Future Work}
\label{sec:conclusion}

We have presented an audio-score piece identification procedure that 
aligns sequences of embeddings learned from sheet music and audio snippets,
and confirmed experimentally, on complex piano music, that our DTW-based matching
strategy performs better than existing alternative methods, for 
all experiment setups and in both search directions.
%
%
The implementation of all evaluated methods is publicly 
available.\footnote{https://github.com/CPJKU/audio\_sheet\_retrieval/tree/eusipco-2021} 

%
%
A central problem of our approach is that the DTW algorithm does not handle
substantial 
structural differences between performances and scores, caused
by, e.g., jumps and repeats.
A number of works \cite{FremereyMC10_RepeatsJumps_ISMIR,GrachtenGAW13_AlignStructDiff_ISMIR,ShanT20_A2SRepeats_Bootleg_ISMIR}
have explored this theme, however we do not consider the proposed solutions
practical for our applications.
Moreover, the scalability experiment uncovered some serious issues regarding the
execution time.
There are ideas for addressing the linear growth issues, but they will require 
some fundamental changes to the model.
As an example, a potential strategy to overcome the two aforementioned problems 
at once 
is to replace the DTW strategy with fingerprints computed over the embeddings.
This approach could result in features that are robust to jumps and repeats in
scores, as well as permit a more time-efficient search engine.


\bibliographystyle{IEEEtran}
\bibliography{refs}

\end{document}